\documentclass[onecollarge,natbib]{svjour2}
\usepackage{amssymb}
   \usepackage{amsmath}
    \usepackage{epsfig}
     \usepackage{bm}
      \usepackage{pifont}

\bibpunct{[}{]}{;}{n}{}{,} 
\smartqed  
\usepackage{graphicx}
\journalname{Few Body Systems}
\begin{document}

\title{Transverse-Distance Dependent Parton Densities in the Large-$x$ Regime
}


\author{I.O. Cherednikov
}


\institute{I.O. Cherednikov \at
              EDF, Departement Fysica, Universiteit Antwerpen,
Antwerp, B-2020 Belgium \\
Bogoliubov Laboratory of Theoretical Physics, JINR,
RU-141980 Dubna, Russia  \\
              \email{igor.cherednikov@uantwerpen.be}             \\
}

\date{Received: date / Accepted: date}

\maketitle

\begin{abstract}
QCD factorization approach in the field-theoretic description of the semi-inclusive hadronic processes in the large Bjorken $x$ approximation implies extraction of the three-dimensional parton distribution functions as a convolution of a collinear jet function and soft transverse-distance dependent (TDD) function defined as a vacuum average of a partially light-like Wilson loop.
The soft function can be interpreted, therefore, as an element of generalized loop space.  A class of classically conformal-invariant transformations of the elements of this space is generated by the non-local area derivative operator which corresponds to a diffeomorphism in the loop space and determines equations of motion, the latter being associated with the rapidity evolution of the soft TDD functions.  We propose a large-$x$ TDD factorization framework and discuss practical applications of this approach to the phenomenology of the TDDs accessible in future experimental programs at the Jefferson Lab 12 GeV and the Electron-Ion Collider.

\end{abstract}

\section{Introduction}
Three-dimensional structure of the nucleon is one of the main objects of investigation in several ongoing and planned experimental and theoretical projects (see, e.g., \cite{INT,JLab12GeV_1,JLab12GeV_2,JLab12GeV_3,JLab12GeV_4,TMD_Pheno_1,TMD_Pheno_2} and Refs. therein). 3D-intrinsic nucleon degrees of freedom become ``visible'' in the semi-inclusive hadronic reactions, which forces us to go beyond the collinear approximation where the well-established QCD factorization methodology applies. In fully inclusive deep inelastic scattering processes, the infinite-momentum framework enables the separation out infra-red and collinear singularities by means of accumulating them into the non-perturbative parton distribution functions (PDFs), as distinct from the hard perturbatively calculable terms. Hence, while the deep inelastic electron-nucleon scattering reactions provides us with the information about the one-dimensional structure of the nucleon,  in semi-inclusive processes (such as, e.g., Drell-Yan and semi-inclusive deep inelastic scattering) one has to take into account as well the intrinsic transverse degrees of freedom of the partons. Transverse structure of the nucleon is encoded in the transverse-momentum dependent PDFs what finalizes the three-dimensional picture in the momentum representation \cite{INT}.

An interesting opportunity to unravel the $3D$ PDFs at large Bjorken $x$ is provided by the forthcoming energy upgrade from 6 to 12 GeV to CEBAF at Jefferson Lab \cite{JLab12GeV_1,JLab12GeV_2}.  Given that CEBAF is a fixed-target facility, this upgrade will enable probing the region $0.1 < x < 0.7$, where valence quarks prevail. On the other hand, smaller $x$ for the similar  $Q^2$ can be reached at the planned Electron-Ion Collider, which will explore the nucleon's sea content as well \cite{INT}. In total, the kinematic range of both experiments is expected to be about $10^{-3}  < x < 1$ and $2$ GeV$^2 < Q^2 < 100$ GeV$^2$. This coverage will allow one to make precision tests of the TMD factorization and compare various methods of it, as well as to look for important relations between nuclear and high-energy phenomenology \cite{LargeX_Accardi}. For original discussion and detailed exposition, see \cite{JLab12GeV_1,JLab12GeV_2} and Refs. therein.

\section{Large-$x$ factorization approach}

The transverse-momentum dependent PDFs can be introduced by means of different methodology  (see, e.g., Refs. \cite{TMD_1,TMD_2,New_TMD_Col_1,New_TMD_Col_2,CS_1,CS_2,CS_3,New_TMD_Feng,New_TMD_SCET_1,New_TMD_SCET_2}). Having in mind the kinematical set up of the above-mentioned experiments, we will discuss the generic $3D$-correlation functions in the large-$x$ limit which is easier to analyze within an appropriate factorization scheme. We will show, moreover, how the theory of the large-$x$ PDFs can profit from the study of properties of the generalized loop space.

We start from a generic gauge-invariant {\it transverse-distance dependent} (TDD) correlation function defined as a Fourier transform of the transverse-momentum dependent hadronic matrix element
${\cal F} \left(x, {\bm k}_\perp; P^+, n^-, \mu^2 \right)$
\begin{eqnarray}
& & {\cal F} \left(x, {\bm b}_\perp; P^+, n^-, \mu^2 \right)
=  \int\! d^2 k_\perp \  {\rm e}^{-ik_\perp \cdot b_\perp} {\cal F} \left(x, {\bm k}_\perp; P^+, n^-, \mu^2 \right) =  \nonumber \\
& &   \int\! \frac{d z^-}{2\pi} \  {\rm e}^{-ik^+ z^-} \ \left\langle
              P \ | \bar \psi (z^-,  \bm{b}_\perp)
              {\cal U}_{n^-}^\dagger[z^-,  \bm{b}_\perp;
   \infty^-,  \bm{b}_\perp] {\cal U}_{\bm l}^\dagger[\infty^-,  {\bm b}_\perp;
   \infty^-,  {\infty}_\perp]  \right.   \\
   && \left.
\cdot \
    {\cal U}_{\bm l}[\infty^-,  {\infty}_\perp;
   \infty^-, \bm{0}_\perp]
   {\cal U}_{n^-}[\infty^-, \bm{0}_\perp; 0^-,\bm{0}_\perp]
   \psi (0^-,\bm{0}_\perp) | \ P
   \right\rangle \nonumber \ ,
\label{eq:TDD_general}
\end{eqnarray}
which is supposed to contain the information about quark distribution in the longitudinal one-dimensional momentum space and two-dimensional impact-parameter coordinate space.
Generic semi-infinite Wilson line evaluated along a certain four-vector $w_\mu$ are defined as
\begin{equation}
{\cal U}_w[\infty ; z]
\equiv {}
 {\cal U}_\gamma = {\cal P} \exp \left[
                      - i g \int_0^\infty d\tau \ w_{\mu} \
                      {\cal A}^{\mu} (z + w \tau)
                \right] \ ,
\end{equation}
where the vector $w_\mu$ parametrizes the path $\gamma : \  w_\mu \sigma$, $\sigma \in [0, \infty]$, the latter containing, in general, light-like, longitudinal non-light-like, and transversal parts \cite{TMD_1,TMD_2}.
Large-$x$ factorization scheme for the gauge-invariant integrated PDFs has been proposed and developed in Ref. \cite{LargeX_KM}. Here we generalize this method  to include the $3D$-PDF, Eq. (\ref{eq:TDD_general}), see Ref. \cite{ChMTvdV_2013}.

First we notice that at the large-$x$, the struck quarks moves as fast as the parent nucleon, that is, in the infinite momentum frame, the soft-gluon contribution is factorized into the eikonal operators \cite{Eikonal}. Thus we re-write Eq. (\ref{eq:TDD_general}) as
\begin{eqnarray}
  {\cal F} \left(x, {\bm b}_\perp; P^+, n^-, \mu^2 \right)
 & = &   \int\! \frac{d z^-}{2\pi} \  {\rm e}^{-ik^+ z^-} \ \left\langle
              P \ | \bar \psi (z)  \  {\cal U}_P [\infty; z] \ {\cal U}_P^\dag [\infty; z] \
              {\cal U}_{n^-}^\dagger[z; \infty]  \right.   \\
   && \left.
\cdot\
   {\cal U}_{n^-}[\infty; 0] \  {\cal U}_P [0; \infty] \   {\cal U}_P^\dag [0; \infty] \
   \psi (0) | \ P
   \right\rangle \nonumber \ ,
\label{eq:TDD_eikonal_0}
\end{eqnarray}
where $z = (0^+, z^-, {\bm b_\perp})$. For the sake of simplicity, we work in what follows in covariant gauge, so the transverse segments of the path are omitted. The eikonal approximation assumes that
very fast quark having the momentum $k_\mu$ can be considered as a classical particle moving parallel to the nucleon momentum $P$, so that instead of the quark fields we use
\begin{equation}
\Psi_{\rm jet} (0) = {\cal U}_P^\dag [0; \infty] \ \psi (0)
\  , \  \bar \Psi_{\rm jet}  (z) = \bar \psi (z) \  {\cal U}_P [\infty; z]  \ ,
\label{eq:jet_WF}
\end{equation}
where the fields $\bar \Psi_{\rm jet}, \Psi_{\rm jet}$ stand for the incoming-collinear jets in initial and final states \cite{Eikonal,LargeX_KM}. Before going over to the large-$x$ approximation, we insert in Eq. (\ref{eq:TDD_eikonal_0}) two full sets of intermediate states
\begin{eqnarray}
 {\cal F} \left(x, {\bm b}_\perp; P^+, n^-, \mu^2 \right)
& & =   \sum_{q} \sum_{q'} \int\! \frac{d z^-}{2\pi} \  {\rm e}^{-ik^+ z^-} \ \left\langle
              P \ | \bar \psi (z)  \  {\cal U}_P [\infty; z] \ | \  q \right\rangle  \left\langle q \  |  {\cal U}_P^\dag [\infty; z] \
              {\cal U}_{n^-}^\dagger[z; \infty]  \right.   \\
   && \left.
\cdot\
   {\cal U}_{n^-}[\infty; 0] \  {\cal U}_P [0; \infty] \  | \  q'  \right\rangle  \left\langle q' \  |  {\cal U}_P^\dag [0; \infty] \
   \psi (0) | \ P
   \right\rangle \nonumber \ ,
\label{eq:TDD_eikonal}
\end{eqnarray}


Now one observes that the large-$x$ regime implies that the struck quark takes almost all momentum of the parent nucleon:
\begin{equation}
k_\mu \approx P_\mu \ .
\end{equation}
Given that in the infinite-momentum frame the transverse component of the nucleon momentum vanishes, the transverse momentum of the incoming quark $\bm k_\perp$ is acquired completely due to the interactions with gluons. The following properties of the large-$x$ regime will be used from now on:

\begin{enumerate}

\item All real radiation can only by soft, that is the intermediate states in Eq. (\ref{eq:TDD_eikonal}) carry zero momenta in the leading approximation:
\begin{equation}
q_\mu, q'_\mu \sim (1- x) P_\mu
\end{equation}

\item Quark radiation is negligible in the leading-twist; virtual gluons can be either soft or collinear, collinear gluons can only be virtual;

\item Rapidity divergences (known also as ``light-cone singularites'') originate only from the soft contributions. This important observation can be justified as follows: given that the rapidity divergence take place in the soft region of the integration over momenta,
that is at small virtual gluon momenta ${\kappa}^+ \to 0$,  one concludes that the minus-infinite rapidity region is responsible for their existence, whose gluons move parallel to the outgoing jet, not incoming-collinear jet, where the rapidity is positive;

\item As distinct from the collinear PDFs, real contributions are ultraviolet-finite due to the transverse distance $\bm b_\perp$ acting as a large-momentum cutoff. They may, however, contain rapidity singularities and exhibit non-trivial $x$- and $\bm b_\perp$-dependence.

\end{enumerate}

It follows immediately from the property (1) that the leading contribution to the large-$x$ TDD (\ref{eq:TDD_eikonal}) is given by the vacuum intermediate state $| 0 \rangle = | q,q' = 0 \rangle$:
 \begin{eqnarray}
  {\cal F} \left(x, {\bm b}_\perp; P^+, n^-, \mu^2 \right)|_{x \to 1}
& = &    \int\! \frac{d z^-}{2\pi} \  {\rm e}^{-ik^+ z^-} \ \left\langle
              P \ | \bar \psi (z)  \  {\cal U}_P [\infty; z] \ | \  0 \right\rangle \left\langle 0 \  |  {\cal U}_P^\dag [0; \infty] \
   \psi (0) | \ P
   \right\rangle
       \\
   & &
   \cdot\
     \left\langle 0 \  |  {\cal U}_P^\dag [\infty; z] \
              {\cal U}_{n^-}^\dagger[z; \infty]   {\cal U}_{n^-}[\infty; 0] \  {\cal U}_P [0; \infty] \  | \  0  \right\rangle  \nonumber \ .
\label{eq:TDD_eikonal_2}
\end{eqnarray}
Using Eq. (\ref{eq:jet_WF}), we write
 \begin{eqnarray}
 & & {\cal F} \left(x, {\bm b}_\perp; P^+, n^-, \mu^2 \right)|_{x \to 1}
 =   \int\! \frac{d z^-}{2\pi} \  {\rm e}^{-ik^+ z^-} \ \left\langle
              P \ | \bar \Psi_{\rm in-jet}  (z)   \ | \  0 \right\rangle \left\langle 0 \  |  \
   \Psi_{\rm in-jet}  (0) | \ P
   \right\rangle
       \\
   & &
   \cdot\
     \left\langle 0 \  |  {\cal U}_P^\dag [\infty; z] \
              {\cal U}_{n^-}^\dagger[z; \infty]   {\cal U}_{n^-}[\infty; 0] \  {\cal U}_P [\infty; 0] \  | \  0  \right\rangle =  \nonumber \\
               & &
            |{\cal J}_{\rm in-jet}  (P)|^2  \  \int\! \frac{d z^-}{2\pi} \  {\rm e}^{-i (1- x)P^+ z^-}  \cdot   \left\langle 0 \  | \   {\cal U}_P^\dag [z ; \infty]
              {\cal U}_{n^-}^\dagger[z; \infty]   {\cal U}_{n^-}[\infty; 0] \  {\cal U}_P [0; \infty] \  | \  0  \right\rangle \ ,
\label{eq:TDD_eikonal_3}
\end{eqnarray}
where $ {\cal J}_{\rm jet}  (P)$ is the jet matrix element  \cite{Li_WL_1,Li_WL_2} which obeys
\begin{eqnarray}
 & &  \left\langle
            P \ | \bar \Psi_{\rm jet}  (z)   \ | \  0 \right\rangle \left\langle 0 \  |  \
   \Psi_{\rm jet}  (0) | \ P
   \right\rangle
   = \nonumber \\
& &    {\rm e}^{-i P^+ z^-}   \left\langle
              P \ | \bar \Psi_{\rm jet}  (P)   \ | \  0 \right\rangle \left\langle 0 \  |  \
   \Psi_{\rm jet}  (P) | \ P
   \right\rangle
   =
   {\rm e}^{-i P^+ z^-}  \  |{\cal J}_{\rm jet}  (P)|^2 \ .
   \label{eq:TDD_fact_1}
\end{eqnarray}

We have shown, therefore, that the following large-$x$ factorization scheme is valid
\begin{equation}
{\cal F} \left(x, {\bm b}_\perp; P^+, n^-, \mu^2 \right)|_{x \to 1}
=
{\cal H} (\mu^2, P^2) \cdot {\Phi} (x, {\bm b}_\perp; P^+, n^-, \mu^2 ) \ ,
\label{eq:LargeX_factor}
\end{equation}
where the contribution of incoming-collinear partons is accumulated in the $x$-independent  jet function ${\cal H}$, while
the soft function $\Phi$ reads
\begin{equation}
{\Phi} (x, {\bm b}_\perp; P^+, n^-, \mu^2 )
= \int\!dz^- \ {\rm e}^{-i (1-x) P^+ z^-} \  \langle 0 | \ {\cal U}_P^\dag [\infty; z] {\cal U}_{n^-}^\dag[z; \infty] {\cal U}_{n^-} [\infty ; 0] {\cal U}_P [0; \infty] \  | 0 \rangle \ ,
\label{eq:soft_LargeX}
\end{equation}
with two sorts of the Wilson lines: incoming-collinear (off-light-cone, $P^2 \neq 0$), ${\cal U}_P$, and outgoing-collinear (light-like, $(n^-)^2 = 0 $), ${\cal U}_{n^-}$.

The properties $(2-4)$ allow us to formulate
the rapidity and renormalzation-group evolution equations for the factorized TDD (\ref{eq:LargeX_factor}):
\begin{eqnarray}
  & & \mu \frac{d}{d\mu} \ln {\cal F} \left(x, {\bm b}_\perp; P^+, n^-, \mu^2 \right)
  = \mu \frac{d}{d\mu} \ln {\cal H} (\mu^2) + \mu \frac{d}{d\mu} \ln {\Phi} (x, {\bm b}_\perp; P^+, \mu^2)
  \label{eq:TDD_evolution_1}
  \ , \\
  & & P^+ \frac{d}{d P^+} \ln {\cal F} \left(x, {\bm b}_\perp; P^+, n^-, \mu^2 \right)
  = P^+ \frac{d}{d P^+} \ln {\Phi} (x, {\bm b}_\perp; P^+, \mu^2)  \ .
  \label{eq:TDD_evolution_2}
  \end{eqnarray}
  Note that the r.h.s. of Eq. (\ref{eq:TDD_evolution_1}) is  $\bm{b}_\perp$-independent and shows up only single-logarithmic dependence on the rapidity \cite{CS_1,CS_2,CS_3}. The r.h.s. of Eq. (\ref{eq:TDD_evolution_2}) corresponds hence to the Collins-Soper-Sterman kernel ${\cal K}_{\rm CSS}$ \cite{CSS_1,CSS_2}.

The rapidity associated with the plus-component of the momentum $P$ is formally infinite and must be supplied with proper regularization \cite{Li_WL_1,CS_1,CS_2,CS_3}.
Namely, one has
\begin{equation}
 Y_P
 =
 \frac{1}{2} \ \ln \frac{(P^+)^+}{(P^+)^-}
 =
\lim_{\eta_P \to 0}  \ \frac{1}{2} \ \ln \frac{(P \cdot n^-)}{\eta_P} \ ,
\label{eq:rapidity}
\end{equation}
where $\eta_P$ is a rapidity cutoff.

Taking into account that the variation of the scalar product $\delta (P \cdot n^-) = \delta S_P$ corresponds to a conformal transformation of the area restricted by the planar part of the path $\gamma*$ on which the Wilson loop is defined
\begin{equation}
{\cal U}_{\gamma*}
\equiv
{\cal U}_P^\dag [\infty; z] {\cal U}_{n^-}^\dag[z; \infty] {\cal U}_{n^-} [\infty ; 0] {\cal U}_P [0; \infty]  \  ,
\label{eq:U_gamma} \
\end{equation}
that is
\begin{equation}
\gamma* =
P_\mu \sigma \cup n^-_\mu \sigma' \cup n^-_\mu \tau'  \cup P_\mu \tau  Ê\  ,
\label{eq:gamma_star}
\end{equation}
with
$$
 \sigma \in [-\infty; 0] \ , \ \sigma' \in [0; \infty] \ , \ \tau' \in [\infty; 0] \ , \  \tau \in [0; \infty] \ .
$$

Eq. (\ref{eq:rapidity}), therefore, implies the simple relationship between rapidity and area logarithmic derivatives:
$$
P^+ \frac{d}{d P^+} = \frac{d }{d \ln S_P} \sim \frac{d }{d Y_P} \ .
\label{eq:nonloc_area}
$$
Hence the rapidity evolution of the soft function (\ref{eq:soft_LargeX}) can be related to the area transformation law of a certain class of the paths.
In the next Section we demonstrate that the non-local classically conformal path variations can be introduced in terms of the so-called Fr\'echet derivative associated to a diffeomorphism in generalised loop space, which makes the whole approach mathematically consistent.

\section{Equations of motion in the loop space and rapidity evolution}	

Generalized loop space consists of the distributional version of closed paths, in the sense that the integration over the path ($\{\gamma_i\}$) is assumed in the base manifold, with an equivalence relation defined in terms of the holonomy of the gauge connection at the base point of the closed paths (see, e.g., Ref. \cite{General_LS} and Refs. therein). This holonomy is given by the operator-valued path-dependent exponential
\begin{equation}
	{\cal U}_{\gamma} = {\cal P} \ \exp{\left[ - ig \oint_{\gamma}\! {\cal A}_\mu(z)\ dz^\mu \right]} \ ,
	\label{eq:paralleltransporter_0}
\end{equation}
where ${\cal A} = A^a t^a$ are the elements of the Lie algebra of the gauge group $SU(N_c)$ and $\{\gamma_i\}$ are elements of generalized loop space. Generic first-order {Wilson loop} ${\cal W}_\gamma^1$ is defined then as a vacuum expectation value of the former:
\begin{equation}
{\cal W}_{\gamma}^1
=
\Big\langle 0 \Big| \frac{1}{N_c} {\rm Tr} \ {\cal U}_\gamma \Big| 0 \Big\rangle \  .
\end{equation}
$n-$th order Wilson loop then reads
\begin{equation}
{\cal W}_{\gamma_1, ... \gamma_n}^{n}
  =
\Big \langle 0 \Big| {\cal T} \frac{1}{N_c} {\rm Tr}\ {\cal U}_{\gamma_1}\cdot \cdot \cdot \frac{1}{N_c}{\rm Tr}\ {\cal U}_{\gamma_n}  \Big| 0 \Big\rangle \ .
   \label{eq:WL_definition}
\end{equation}
Information on the geometrical and dynamical properties of the loop space, which can include cusped light-like Wilson exponentials, will help us to understand fundamental properties of the renormalization  various gauge-invariant quantum correlation functions, in particular, their renormalization behaviour and evolution \cite{MM_WL_1,MM_WL_2,CS_1,CS_2,CS_3}.

In the generalized loop space one can introduce local (i.e., defined at a given point in the space-time, not necessary lying on the path) differential operators such as the path- and area-derivative
used by Makeenko and Migdal to derive their non-perturbative equations \cite{MM_WL_1,MM_WL_2,WL_Renorm_1,WL_Renorm_2}:
\begin{equation}
 	\partial_x^\nu \ \frac{\delta}{\delta \sigma_{\mu\nu} (x)} \ {\cal W}^1_\gamma
 	=
	 N_c g^2 \ \oint_{\Gamma} \ dz^\mu \ \delta^{(4)} (x - z) {\cal W}^2_{\gamma_{xz} \gamma_{zx}}  \ .
 \label{eq:MM_general}
\end{equation}
However, some physically interesting transformations of the cusped (partially) light-like paths form a special class of variations which is not accessible directly within the Makeenko-Migdal approach.
Recently we have shown that the logarithmic Fr\'echet derivative \cite{General_LS}
		\begin{equation}
			D_V \  {\cal U}_\gamma
			=
			{\cal U}_\gamma \cdot \int_0^1\! dt \
			{\cal U}_{\gamma^t} \cdot \
			{\cal F}_{\mu\nu} (t) \  \left[V^\mu(t) \wedge \dot{\gamma}^\nu(t) \right] \cdot {\cal U}_{\gamma^t}^{-1} \ ,
			\label{eq:frechetder1}
		\end{equation}
		where $ {\cal F}_{\mu\nu}$ is the gauge-field strength and $\gamma (t)$ stands for the parameterisation of the path,
being applied to a generic operator-valued parallel transporter (Eq. (\ref{eq:paralleltransporter_0}))  along the segment of the path $\gamma$ from the point $0$ to $t$ is equivalent (at least, in the leading order of perturbative expansion) to the non-local area derivative (\ref{eq:nonloc_area}), Ref. \cite{ChMVdV_2012,ChMVdV_2013,MT_2013,ChM_2014_1}. In this case we must assume that the vector field $V$, associated with the diffeomorphism flow, is parallel to $P^+$, and the path is parameterized according to Eq. (\ref{eq:gamma_star}). Then one obtains
\begin{equation}
 \mu \frac{d}{d\mu} \  D_P \  {\cal U}_{\gamma*}  =
 \mu \frac{d}{d\mu} \ \frac{d }{d Y_P}  \  {\cal U}_{\gamma*}
 =
- 2 \Gamma_{\rm cusp} \ ,
\label{eq:final}
\end{equation}
where $\Gamma_{\rm cusp}$ is the light-cone cusp anomalous dimension \cite{KR87,WL_LC_rect_1,WL_LC_rect_2,WL_LC_rect_3} and the factor 2 corresponds to the total number of cusps in two parallel planes.

To conclude, we have proposed the factorization of the three-dimensional transverse-distance dependent parton densities valid in the large Bjorken $x$ approximation. We argue that the calculation of the evolution kernels (\ref{eq:TDD_evolution_2}) simplifies by making use of the geometrical properties of the generalized loop space. In particular,
Eq. (\ref{eq:final}) allows us to express the rapidity evolution of a given element of the loop space in terms of the logarithmic Fr\'echet derivative, which can be then applied to obtain the solution of the combined rapidity-ultraviolet evolution equations (\ref{eq:TDD_evolution_2}). The result will be reported elsewhere.

\section{Acknowledgements}
I thank Tom Mertens, Frederik Van der Veken and Pieter Taels for long-term fruitful collaboration and inspiring discussions.

\end{document}